\documentclass[letterpaper,12pt]{DecadalWhitePaper}
\usepackage[letterpaper,top=1in, bottom=1in, left=1in, right=1in]{geometry}
\usepackage[utf8x]{inputenc}
\usepackage[T1]{fontenc}
\usepackage[caption=false,font=footnotesize]{subfig}
\usepackage[hyphens]{url}
\usepackage[colorlinks=true, bookmarks=true, hidelinks]{hyperref}
\usepackage[numbers,sort&compress]{natbib}
\usepackage{amssymb}
\usepackage{amsmath}
\usepackage[english]{babel}
\usepackage{graphicx}
\usepackage[usenames,dvipsnames]{color}
\usepackage{sidecap}
\usepackage{pdfpages}
\usepackage{wrapfig}
\usepackage{placeins}
\usepackage{cleveref}

\usepackage{array,booktabs,ragged2e}
\usepackage{makecell}

\newcolumntype{R}[1]{>{\RaggedLeft\arraybackslash}p{#1}}

\graphicspath{{figures/}}

\begin{document}

%
\pagenumbering{gobble}
\title{\vspace{-2cm}\centering Trinity: An Air-Shower Imaging Instrument to detect Ultrahigh Energy Neutrinos\vspace{2pt}
    \includegraphics*[width=0.85\textwidth]{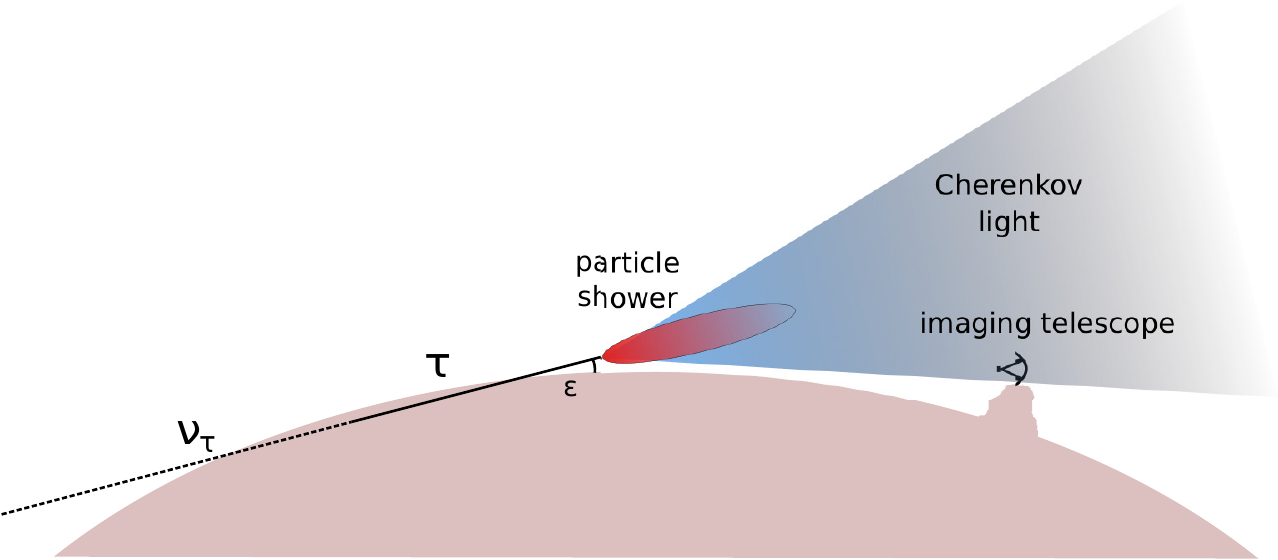} 
}
\abstract{ \vspace{-2ex}
Trinity is a proposed air-shower imaging system optimized for the detection of earth-skimming ultrahigh energy tau neutrinos with energies between $10^7$ GeV and $10^{10}$ GeV. Trinity will pursue three major scientific objectives. 1) It will narrow in on possible source classes responsible for the astrophysical neutrino flux measured by IceCube. 2) It will help find the sources of ultrahigh-energy cosmic rays (UHECR) and understand the composition of UHECR. 3) It will test fundamental neutrino physics at the highest energies.\\
Trinity uses the imaging technique, which is well established and successfully used by the very high-energy gamma-ray community (CTA, H.E.S.S., MAGIC, and VERITAS) and the UHECR community (Telescope Array, Pierre Auger).
}
\maketitle

\noindent\textbf{Consideration area:} Particle Astrophysics and Gravitation, Compact Objects and Energetic Phenomena, Cosmology, Galaxies \vspace{1ex}

\noindent\textbf{Authors:}\\
Prof.\ Anthony M.~Brown, Durham University, United Kingdom\\
Prof.\ Michele Doro, INFN Padova Italy\\
Prof.\ Abe Falcone, Pennsylvania State University\\
Prof.\ Jamie Holder, University of Delaware\\
Dr.\ Eleanor Judd, UC Berkeley, Space Sciences Laboratory \\
Prof.\ Philip Kaaret, University of Iowa\\
Prof.\ Mosè Mariotti, INFN Padova, Italy\\
Prof.\ Kohta Murase, Pennsylvania State University\\ 
Prof.\ A.~Nepomuk Otte, Georgia Institute of Technology\\
Prof.\ Ignacio Taboada, Georgia Institute of Technology\vspace{2ex}

\noindent\textbf{Contact:}\\
A.~Nepomuk Otte, Georgia Institute of Technology \& Center for Relativistic Astrophysics\\
email: otte@gatech.edu

\setcounter{page}{1}
\pagenumbering{arabic}
\section{Science Goals and Objectives: Opening the UHE neutrino band}
\begin{figure}[tbp]
  \centering
  \includegraphics*[angle=0,width=0.9\textwidth]{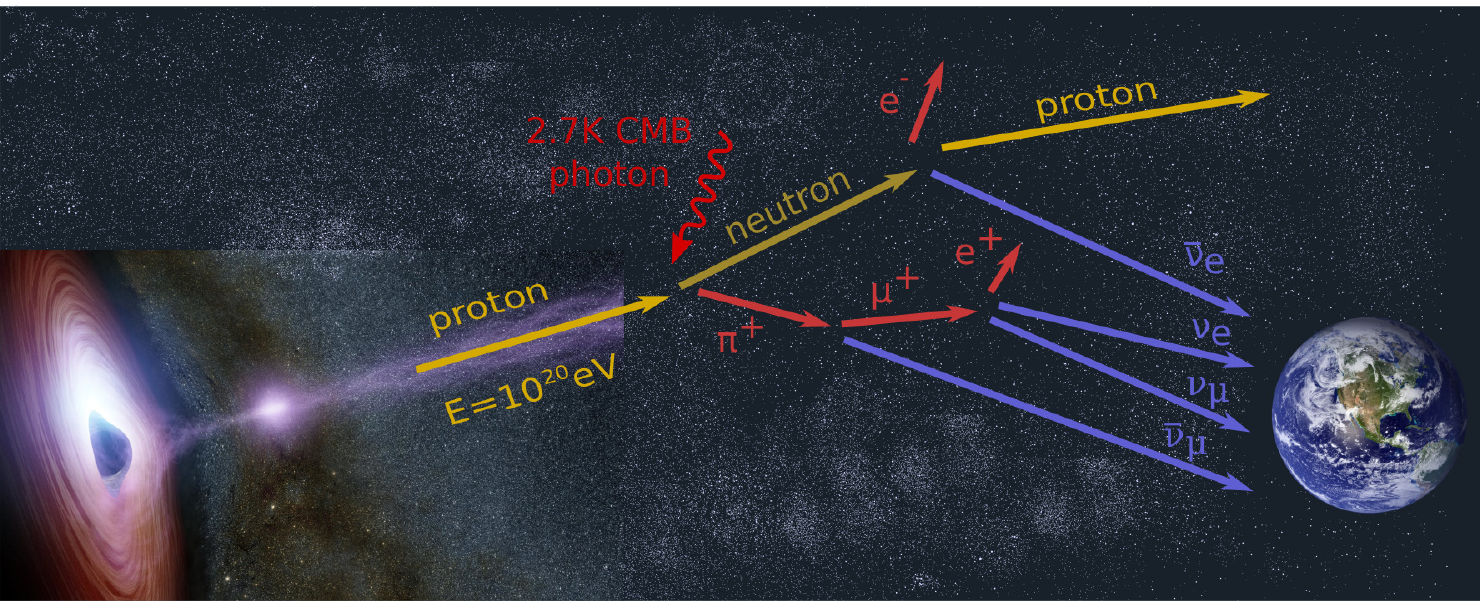}\vspace{2pt}
  \caption{Production of ultrahigh energy neutrinos via the collision of cosmic microwave background photons with ultrahigh energy cosmic-ray protons.}
  \label{fig:EarthSkimming}
\end{figure}


The last ten years have brought us the stunning capabilities of detecting gravitational waves and astrophysical high-energy neutrinos. With these new windows opened to the universe, we witness the dawn of multimessenger astrophysics and its enormous potential. For example follow-up observations in the electromagnetic spectrum of a GW-detected merger of two neutron stars identified them as the origin of short GRBs and places where efficient element synthesis takes place (r-process) \cite{2017ApJ...848L..12A,2017ApJ...848L..13A} (see also a white paper \cite{2019BAAS...51c..38B}). Another example of the potential of multimessenger astrophysics is the recent  evidence that blazars are a source of neutrinos; a conclusion made possible by combining neutrino observations with very-high energy gamma-ray observations  \cite{IceCubeCollaboration2018a,IceCubeCollaboration2018b}. Opening a new window to the universe and combining it with other windows will always result in stunning new views, which are not possible otherwise.

One window that is still closed is the ultrahigh energy (UHE) neutrino window ($>10^{7}$\,GeV), which is adjacent to the high-energy neutrino band where the IceCube team detected an astrophysical neutrino flux. Recent advances in technology and experimental techniques make new UHE instruments feasible, which have sensitivities that tap into expected UHE neutrino fluxes. \emph{Trinity} promises to be one such new UHE neutrino detector.

The IceCube detection of astrophysical neutrinos raises a number of interesting questions that are potentially answered by neutrino observations at higher energies: 
What are the astrophysical sources of these neutrinos \cite{2016PhRvD..94j3006M,2017NatPh..13..232H,2019arXiv190610212M} (see also a white paper \cite{2019arXiv190304334A})?
What is the multimessenger connection among neutrinos, UHE cosmic rays and gamma rays (see \cite{2013PhRvD..88l1301M,2016PhRvD..94j3006M,2018NatPh..14..396F})? 
Are blazars really a source of the IceCube detected neutrinos \citep[e.g.][]{2019arXiv190304447B,2018ApJ...865..124M}? 
What other types of astrophysical neutrino sources are there? 
In order to address these questions, it is particularly important to extend the spectral measurements of the astrophysical neutrino flux to higher energies and to reconstruct the arrival direction of neutrinos with better angular resolution.
Trinity promises to explore the high-energy end of the diffuse neutrino spectrum discovered in IceCube and to reveal their connection with gamma rays and UHE cosmic rays.   

But, hunting down IceCube neutrinos and probing their source physics is not the only science case for UHE neutrino detectors; the long-standing quest to understand the composition and the sources of UHE cosmic rays is another one \cite{2019BAAS...51c..93S,2019arXiv190306714A}. 
The connection between neutrinos and cosmic rays is made when UHE cosmic rays interact with cosmic background radiation \cite{1969PhLB...28..423B}. The number of neutrinos produced in these interactions depends on the composition of the cosmic rays, redshift evolution of the sources, and the maximum energy of UHE cosmic rays\,\citep[e.g.][]{Kotera2010}.  
While recent measurements with UHE cosmic-ray experiments favor a heavier composition\,\citep[e.g.][]{Kampert2012,ThePierreAugerCollaboration2016,Yushkov2017,Aab2017a,Aab2017}, which leads to pessimistic fluxes of cosmogenic neutrinos \citep[e.g.][]{2019JCAP...01..002A,2019ApJ...872..110B,2018arXiv181210289Z}, the details of the neutrino-flux predictions depend on the galactic-extragalactic transition model. For example, higher fluxes are predicted if an extragalactic component emerges around the second knee at $\sim10^{17}$~eV \cite{2005PhRvD..72b3001A,2009APh....31..201T,2016PhRvD..94d3008L,2018NatPh..14..396F}.  

The benchmark upper bound flux of astrophysical neutrinos from the sources of UHE nuclei is a few $10^{-9}$\,GeV\,cm$^{-2}$\,s$^{-1}$\,sr$^{-1}$, which is within Trinity's reach \cite{2010PhRvD..81l3001M}.
The energy range covered by Trinity is critical to study the galactic-extragalactic transition in the cosmic-ray spectrum and the transition from source neutrinos to cosmogenic neutrinos. 
More importantly, UHE neutrino observations will provide a unique handle to constrain source models of UHE cosmic rays by combining them in a multimessenger approach with GeV-TeV gamma-ray and UHE cosmic-ray data. 
Furthermore, a combination of UHE neutrino measurements with proton measurements obtained with next-generation UHE cosmic-ray observatories (e.g. AugerPrime) would allow one to simultaneously constrain the UHE proton fraction and source evolution \cite{2019arXiv190101899V}.

The third science case is tests of neutrino physics at the highest energies, which could potentially hint at new physics beyond the standard model
\citep[see white papers][]{GRANDCollaboration2018,2019arXiv190304333A,2019arXiv190700991B}. 
One approach would be to compare neutrino fluxes measured with experiments that are sensitive to different neutrino flavors. In that context it is noteworthy that experiments geared toward measuring extraterrestrial neutrinos have already provided numerous important contributions to neutrino physics \citep[e.g.][]{Cleveland1998,Fukuda1998,Ahmad2001,Ahmad2002,IceCubeCollaboration2017,Aartsen2018a}.
UHE neutrinos can be used as an important probe of beyond Standard Model physics, neutrino cross sections~\citep[e.g.][]{Klein2013,2019arXiv190304333A}, and superheavy dark matter~\citep[e.g.][]{Ryabov2006,2012JCAP...10..043M,2018PhRvD..98h3016K,2019JCAP...05..051B,2019arXiv190304333A}. 
Adding to the discovery potential for new physics is the detection of
neutrino candidates with ANITA, which have signatures expected from air showers but seem to contradict the current understanding of neutrino physics \citep[e.g.,][]{Gorham2018a,Gorham2019,2018arXiv180909615F,2019PhRvD..99d3009C,2019PhLB..790..578A,2019arXiv190700991B}.

\begin{SCfigure}[1.0][t]
  \centering
    \includegraphics*[width=0.55\columnwidth]{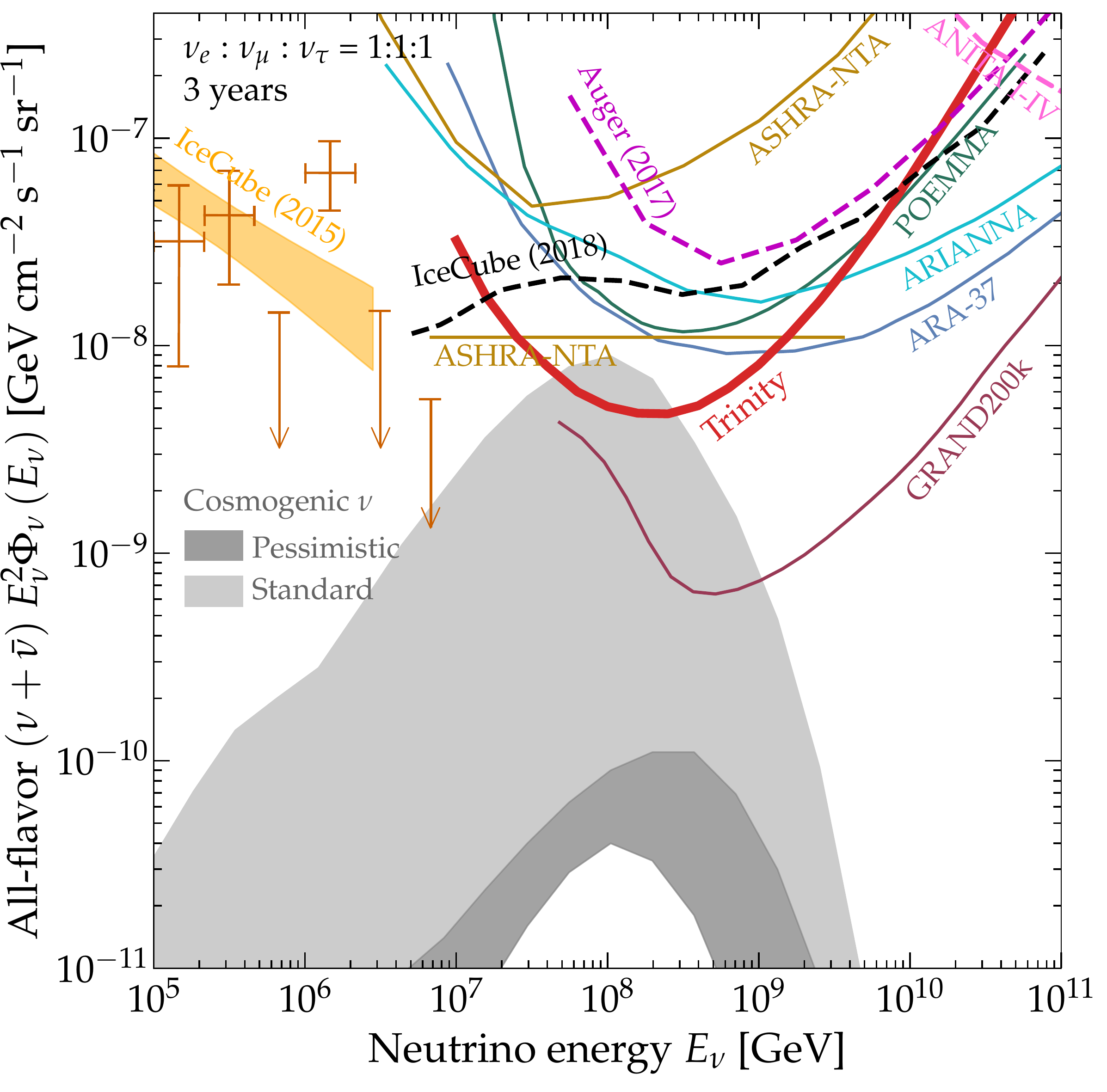} 
  \caption{Sensitivity of \emph{Trinity} and other proposed experiments with flux measurements and recent limits. The sensitivities are quoted for an all flavour neutrino flux and three years of observation. For \emph{Trinity} a 20\% duty cycle is assumed. Figure from \cite{Otte2019a}.}
  \label{fig:TrinitySens}
\end{SCfigure}

Addressing any of the above question requires a differential sensitivity in the UHE band of a few $10^{-9}$\,GeV\,cm$^{-2}$\,s$^{-1}$\,sr$^{-1}$. \emph{Trinity} will reach this sensitivity (see Figure \ref{fig:TrinitySens}).

\section{Technical Overview}

\paragraph{\emph{Trinity} - architecture} \emph{Trinity} is a system of air-shower imaging telescopes that is installed at an altitude of 2\,km - 3\,km. All telescopes of the system point at the horizon, cover the entire azimuthal range, and take images of particle showers, which are the result of tau neutrinos interacting inside the Earth. In the baseline design, six telescopes, each with a $60^\circ$ field of view, monitor $360^\circ$ in azimuth, which results in the sensitivity shown in figure \ref{fig:TrinitySens}. Such an azimuthal coverage and sensitivity can be achieved by either deploying all telescopes at one site or distributing the telescopes to different sites.

\paragraph{Key performance requirement: Imaging of air showers} \emph{Trinity}'s task is to image particle showers in the atmosphere (air showers). The concept is illustrated on the cover page. A tau neutrino interacts inside the Earth and produces a tau, which emerges from the ground and starts a particle shower in the lower atmosphere. Electrons and muons produced in electron neutrino and muon neutrino interactions, respectively, seldom generate air showers. The telescopes of \emph{Trinity} take an image of the air shower by collecting Cherenkov and fluorescence light that is produced by the shower particles. Based on the characteristics of the image, the arrival direction and the energy of the neutrino can be reconstructed. 

Air-shower imaging is an established technique, which is used with imaging atmospheric Cherenkov telescopes (IACTs) like H.E.S.S.\ \cite{Aharonian2006a}, MAGIC \cite{Aleksic2016}, and VERITAS \cite{Holder2016}. It is also applied by the ultrahigh-energy cosmic ray (UHECR) community in the Pierre Auger Observatory \cite{ThePierreAugerCollaboration2017} and Telescope Array \cite{Tokuno2012}. The imaging technique has excellent discriminatory power to distinguish between particle-shower images and spuriously triggered events, which result from noise and fluctuations in ambient photon backgrounds.

Imaging air showers from earth-skimming neutrinos has been proposed in the late 90s \cite{Fargion1999}. The most recent limits on the UHE neutrino flux where the imaging technique has been applied come from the MAGIC collaboration \cite{Ahnen2018a}. That study underlines the intrinsic robustness of the imaging technique and its general feasibility to pursue UHE neutrino physics. But the MAGIC limits are not competitive because they are the result of a limited, 30\,hour long exposure. Another limiting factor is the restricted angular acceptance of MAGIC ($3^\circ$). The design and operation of \emph{Trinity}, on the other hand, is optimized for the detection of earth-skimming tau neutrinos and does not have these limitations. 

\emph{Trinity} achieves a competitive sensitivity by being able to detect a sufficient amount of light from an air shower that is induced by a neutrino with an energy as low as $10^7$\,GeV and that can be as far away as 200\,km. The expected three year, all flavor sensitivity of \emph{Trinity} is depicted in figure \ref{fig:TrinitySens}. In achieving the projected sensitivity, \emph{Trinity} will perform similarly to other proposed UHE detectors and observe complementary parts of the sky.

\subsection{Instrument Requirements and Implementation}

The instrument requirements for \emph{Trinity} are the results of a detailed study \cite{Otte2019a} and the experience gained by working with Cherenkov telescopes. The key parameters of the system are summarized in table \ref{tab:InstPars}. 

\begin{table}[htb]
\centering
\caption{Main system requirements for \emph{Trinity}}
\label{tab:InstPars}
\begin{tabular}{ll}
\textbf{Parameter}                 & \textbf{Value}  \\ \hline
Duty cycle & 20\% (1800 hours per year) \\
Effective light collection surface & $>10$\,m$^2$        \\
Vertical field of view             & 5$^\circ$       \\
Azimuthal field of view           & 360$^\circ$     \\
Pixel size                         & $0.3^\circ$     \\
Sensitive wavelength range         & 340\,nm - 900\,nm \\
Readout sampling speed             & 100 mega-samples per second\\
Readout amplitude resolution       & 1 photoelectron \\
Dynamic Range                      & 1000 photoelectrons 
\end{tabular}
\end{table}

The design of the optics and the readout of \emph{Trinity} follows that of any air-shower imaging system. Each telescope of the system consists of a mirror that collects light from the air shower and projects it onto a camera, which takes an image of the air shower. The pixels of the camera are fast, single photon detecting light sensors -- silicon photomultipliers (SiPMs) in the case of \emph{Trinity}. All pixels are connected to a common readout system, which determines when the camera has recorded a useful image and saves the image to disk.

Adding to the instrument requirements is the desire to keep operating costs minimal. That is accomplished by operating remotely and minimizing the need for maintenance. That air-shower imaging systems can be run remotely is demonstrated by the FACT team, which has been operating a Cherenkov telescope for almost ten years \cite{Anderhub2013}. The need for maintenance is minimized by implementing ruggedized mirror technologies that have been developed by the Cherenkov telescope community and are being used in the Cherenkov Telescope Array (CTA). The reflectivity of these mirrors degrades only a few percent each year, even though they are not protected from the elements. Also minimizing the need for maintenance is the use of SiPMs, which are known to not age even after several years of use unlike classical photomultipliers \cite{Anderhub2013}.

\begin{SCfigure}[1.0][t]
  \includegraphics*[width=0.68\columnwidth]{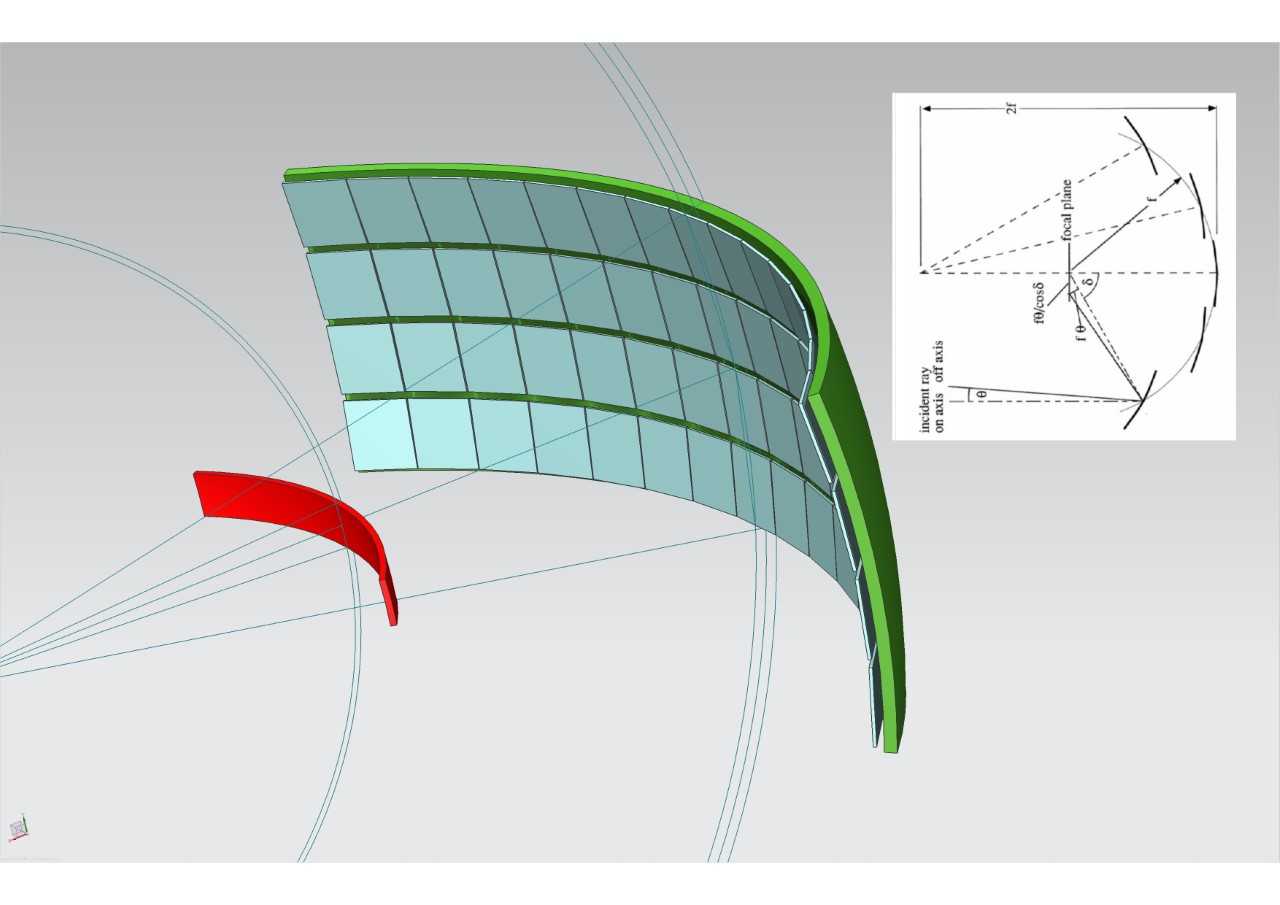} 
  \caption{Optical design of one \emph{Trinity} telescope. The focal length is 5.6\,m. The light collecting surface of $4$\,m$\times17$\,m is tessellated with $1$\,m$\times1$\,m sized mirror elements. Each of the 3,300 pixel in the focal plane (red) has a size of 19\,mm$\times$19\,mm and accepts light from a $4\,$m$\times4$\,m large portion of the light collecting surface. }
  \label{fig:optics}
\end{SCfigure}

\paragraph{Optics}
\emph{Trinity} is a system of several telescopes. In order to achieve a large field of view in azimuth and thus minimize the number of telescopes, it is planned to implement the optics developed for MACHETE \cite{Cortina2016}. That optics delivers a $5^\circ$ field of view in the vertical and a $60^\circ$ field of view in azimuth. Figure \ref{fig:optics} shows a scaled version of MACHETE that meets the requirements for \emph{Trinity}. It achieves an optical point spread function of $0.3^\circ$ and provides a 16\,m$^2$ light collection surface everywhere in the field of view.

\paragraph{Camera}
The camera of a \emph{Trinity} telescope is equipped with 3,300 silicon photomultipliers (SiPMs). SiPMs perfectly match the Cherenkov spectrum, which is heavily attenuated in the blue and peaks in the red \cite{Otte2019a}. SiPMs offer a variety of additional advantages like no aging, bias voltages of less than 100\,V, no damage when exposed to daylight even when under bias. For the amplification of the SiPM signals and the adjustment of the SiPM bias voltages it is planned to use an Application Specific Integrated Circuit (ASIC), called MUSIC \cite{Gomez2016}. MUSIC was specifically developed to be used with SiPMs in air-shower imaging applications.

The readout system has to be capable of sampling the SiPM signals with a speed of at least 100 megasamples per second or faster and a resolution of at least 8 bit. Such a slow sampling speed is sufficient because Cherenkov photons arrive over tens of nanoseconds when an air shower is viewed under an angle of several degrees \cite{Otte2019a}. Commercially available readout systems that meet these requirements are readily available. 

Presently, it is planned to adopt the AGET readout for \emph{Trinity}, a switch-capacitor-array based system \cite{Pollacco2018}. A Cherenkov telescope camera that uses SiPMs and the AGET system is presently under development for a balloon-borne Cherenkov telescope \cite{Otte2019b}. It is planned to use the same system for \emph{Trinity} with minor adjustments like scaling the pixel size to fit the larger plate scale of the \emph{Trinity} optics. Each \emph{Trinity} pixel is composed of a non-imaging light concentrator made from UV transparent acrylic with an entrance area of 19\,mm$\times$19\,mm that is coupled to a 9\,mm$\times$9\,mm SiPM.

\paragraph{Data volume:}
The expected data rate is dominated by accidental triggers, which are caused by fluctuations in the ambient photon background at the trigger threshold. The trigger threshold thus defines the data rate and is adjusted such that the trigger rate is a few events per second. For the sensitivity calculation shown in figure \ref{fig:TrinitySens} the assumed trigger threshold results in an accidental trigger rate of one event per second. For the estimated data volume an accidental trigger rate of ten events per second is assumed. The rate of cosmic-rays will be much lower. Each triggered event generates about 1\,kB of data resulting in about 30\,MB of data in a typical observing night. A larger amount of data will be accumulated by the slow control, which monitors the health of the instrument and by the surveillance of the vicinity around the telescope. The generated data volume is small.

\subsection{Calibration}

\paragraph{Camera calibration}
During prototyping and commissioning, the pixel linearity, angular response and pixel gain of \emph{Trinity}'s camera must be characterised. However, due to variations in ambient conditions such as temperature or ambient background-photon intensities, this initial characterisation needs to be routinely updated. Such a monitoring of \emph{Trinity's} camera performance will be achieved with compact, on-structure calibration systems. Based on CTA prototypes \citep[e.g.][]{brown2015}, such a system will be capable of emitting pulses of different wavelengths with durations of 4 nanoseconds, across a large dynamic range (from 0.1 to 1000 photoelectrons).

\paragraph{Atmospheric calibration}
Reconstructing the neutrino's energy from the recorded air-shower image requires adequate monitoring of the extinction in the atmosphere. Atmospheric extinction measurements over large distances can be achieved through a number of complimentary methods, such as continually monitoring the intensity of stars close to the horizon and monitoring the return and reflected signal from a pulsed laser fired into the direction of the air shower (a lidar). In addition, new unmanned aircraft technology allow us to fly calibrated reference light sources at a range of distances from \emph{Trinity} to periodically cross-calibrate these complimentary methods \citep[e.g.][]{2018APh....97...69B}). We note however, that a complete calibration concept still needs to be developed.

\subsection{Site and Infrastructure Requirements}
\paragraph{Site} Air showers initiated by earth-skimming neutrinos typically develop in the lower atmosphere \cite{Otte2019a}. To be able to record these air showers, which can develop in a distance of 200\,km and whose light is heavily absorbed inside a haze layer, which extends from the ground to an altitude of 1\,km to 2\,km, \emph{Trinity} has to be located above the haze layer. An additional site requirement is that artificial light sources are kept to a minimum inside the telescopes field of view. Less critical are the sky conditions; clouds at high altitudes, \emph{i.e.}\ several kilometers above ground do not affect the duty cycle of the experiment. Suitable sites are located for example in Utah. A proper site search has not yet been conducted.

Even though \emph{Trinity} will be remotely operated, infrequent access will be required to visually inspect the system and conduct maintenance.

\paragraph{Buildings} The \emph{Trinity} telescopes will be located in buildings like the fluorescence detectors of the Pierre Auger Observatory or Telescope Array to protect them from the elements but also to prevent the sun from damaging the telescopes during sunset and sunrise.

\paragraph{Power} The power consumption of \emph{Trinity} will be about 4\,kW during observation. The estimate is based on measurements of main components in the signal chain, which amounts to 2.5\,kW. The remaining 1.5\,kW are needed for computing and auxiliary equipment. 

\paragraph{Internet connection} For remote operation and data transfer a remote connection will be required with a minimum speed of 20 Mbps, \emph{e.g.}\ a 4G LTE connection. 

\subsection{Public-Private Partnership}

A public-private partnership is not anticipated for construction and operation.

\section{Technology Drivers}

No new technologies need to be developed for Trinity. The project will however benefit from ongoing developments of mirror technologies and SiPM technologies, which might become available by the time construction starts.


\section{Organization, Partnerships, and Current Status}
\emph{Trinity} is a small enough project to be constructed and run by a collaboration of university groups. A formal collaboration with an appropriate governance structure does not yet exist but will be established when funding will be sought for construction. All authors have stated their interest to participate in the project. They bring to the table extensive experience in the construction and operation of air-shower imaging instruments like CTA, H.E.S.S., MAGIC, the Pierre Auger Observatory, and VERITAS. Some have more than 20 years of experience in the field. The present group has experts for every aspect of the instrument, \emph{i.e.}\ the optical system including mirror fabrication, silicon photomultipliers, frontend electronics, and data acquisition. The majority of the participating groups are based at US research institutes -- Colorado School of Mines; Georgia Tech; UC Berkeley, SSL; Penn State; University of Delaware; and the University of Iowa. Groups participating from outside the US are INFN Padova, Italy and Durham University, UK.

\section{Schedule}

\paragraph{Development}

A five year development cycle is needed to finalize the design of the telescopes, construction of a prototype telescope and construction of the final instrument and infrastructure on site. It is assumed that an appropriate site has been determined before the start of the development cycle.

\paragraph{Year 1 and year 2:}  Design of the optical support structure of the telescope and construction of a prototype with mirrors by the end of year one. In parallel, a prototype camera will be developed. The electronics for the camera already exists but some development work is needed to integrate the SiPMs and the non-imaging light concentrators. The third activity is the development of on-site infrastructure, which includes power, data link, and construction of the building for one telescopes. At the end of year 2 a prototype telescope is installed on site.

\paragraph{Year 3:} Commissioning of the prototype telescope. Revision of the telescope design at the end of year 3. 

\paragraph{Year 4 and year 5:} Start of science operation with the prototype telescope. Successive deployment of the remaining telescopes and integration into science operation.

\paragraph{Operational lifetime:} After deployment of the complete system, the operational lifetime will be three years in order to achieve the sensitivity in figure \ref{fig:TrinitySens}. Continued operations will depend on the scientific results achieved in the first three years. 


\section{Cost Estimate}
The most up-to-date cost estimate for Trinity construction and 3-year operation brings it to a total budget that is under \$7.4M (\$4.5M for construction and $<\$2.9\,$M for 3-year operation). This puts Trinity in the category of small ground-based experiments, which are defined as having expected overall costs (including operations and science) that are less than \$20\,M over an operation period of three years. The cost estimate has been updated in June 2019 for the fiscal year 2020. The estimate is based on quotations where possible and based on experience gained in the construction of the Cherenkov telescopes CTA, MAGIC, VERITAS elsewhere. 

\subsection{Construction}

The costs for constructions are summarized in table \ref{tab:costs} and discussed in the following.

\begin{table}[htb]
\centering
\caption{Cost estimate for the construction of six telescopes.}
\label{tab:costs}
\begin{tabular}{p{6.0cm}>{\RaggedLeft}p{2.0cm}}
\toprule
\thead{Item} & \thead{Cost} \\
\midrule
Camera \& readout  & \$2,000,000        \\
Optics  & \$1,000,000        \\
On-site computing & \$100,000 \\
Infrastructure & \$1,400,000 \\
\multicolumn{1}{r}{\textbf{Total costs of \emph{Trinity}:}} & \textbf{\$4,500,000} \\
\bottomrule
\end{tabular}
\end{table}

\paragraph{Camera and readout} The majority of the components for the camera and readout are commercially available. One exception is the frontend electronics -- signal amplification, SiPM housekeeping, temperature monitoring, single pixel trigger, which requires the development of a custom printed circuit board, which integrates the MUSIC chip. 

The application of MUSIC, SiPMs, and the AGET readout system results in significant cost savings and a per channel costs of \$100. For the whole system, \emph{i.e.}\ 20,000 channels, the cost for cameras will be \$2M.

\begin{SCfigure}[1.0][tbp]
  \centering
  \includegraphics*[angle=0,width=0.5\textwidth]{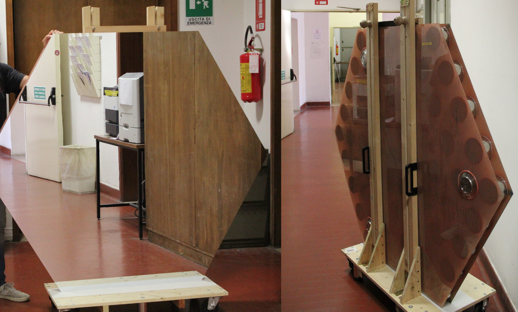}\vspace{2pt}
  \caption{Prototype mirrors produced with the thin glass replica method. The mirror has a diameter from edge to edge of 1.5\,m. The mirrors for \emph{Trinity} will be square with a length of 0.5\,m to 1\,m.}
  \label{fig:mirror}
\end{SCfigure}

\paragraph{Optics} The optical system is divided into the mirrors, the structure that holds the mirrors, and the non-imaging light concentrators, which attach to the SiPMs. The light-collecting surface of the telescope is tesselated with $1$\,m$^2$ mirrors, see figure \ref{fig:optics}. Several technologies exist to produce mirrors of the required size and adequate optical quality. One of them is the thin foil glass replica technology. Figure \ref{fig:mirror} shows a 1.5\,m diameter mirror produced in that technology for CTA. The point spread function of that mirror is $0.05^\circ$, much below the intrinsic $0.3^\circ$ resolution of the MACHETE optics. The cost of the thin film foil glass replica technology is \$2k/m$^2$ or \$600k to complete all \emph{Trinity} telescopes. The mechanical structure to support the mirrors and the camera is estimated to be \$25k per telescope or \$150k for all telescopes. The light concentrators are \$15 each or \$250k for all pixels, which includes the mold to fabricate them. 

\paragraph{On-site computing} Computing is needed to control the system and save the data to disk. A redundant computing architecture is envisioned for on site computing, with a primary and secondary server. Because not much storage is required, \$100k should be sufficient for computing. A detailed cost estimate has not yet been performed for the computing. 

\paragraph{Infrastructure} On-site infrastructure includes buildings to host the telescopes and the computing infrastructure. It also includes fencing, a data link, calibration infrastructure, and power. The infrastructure costs are the least developed in part because a proper site search is still pending. For the time being, a conservative approach is taken to estimate the infrastructure costs. For the estimate we assume a site that is accessible with heavy equipment and has no existing infrastructure. Power (40kWh per day) has to be generated on site either with a wind turbine, solar, or a combination of both with a generator as backup. Costs for power including installation are estimated to be \$200k. The costs of buildings and fences are estimated to be \$1M. Data link and calibration infrastructure are estimated to be \$200k.

\subsection{Operational costs} Expenses to run the experiment are mainly to support graduate students and postdoctoral researchers, who analyze data and maintain the instrument. A small amount of funding will be needed for materials and supplies to maintain the operation of the instrument. Because \emph{Trinity} will be operated remotely, a permanent on-site presence of technical staff is not necessary. The operational costs are summarized in table \ref{tab:opcosts}. 

\begin{table}[htb]
\centering
\caption{Operational costs per year}
\label{tab:opcosts}
\begin{tabular}{p{6cm}>{\RaggedLeft}p{2.0cm}}
\toprule
\thead{Item} & \thead{Cost} \\
\midrule
6 graduate students  & \$400,000        \\
4 Postdoctoral researchers  & \$400,000 \\
Technical support staff  & \$50,000     \\
On-site travel & \$40,000 \\
Hardware maintenance & \$60,000 \\
\multicolumn{1}{r}{\textbf{Total per year:}} & \$950,000 \\
\bottomrule
\end{tabular}
\end{table}

Monthly maintenance visits are necessary to visually inspect the instrument and its vicinity as well as upgrade and repair components of the instrument. These visits will be mostly carried out by faculty, postdoctoral researchers and graduate students. And they will serve as an important part of the training of the next generation of experimental astroparticle physicists.

Decommissioning costs of about \$400k cover the removal of the instruments and support buildings. Unusual decommissioning costs are not expected.

\newpage
\renewcommand\thesection{}
\section{Bibliography \& References Cited}
\renewcommand\refname{}
\bibliography{references}

\begin{thebibliography}{10}
\providecommand{\url}[1]{\texttt{#1}}
\providecommand{\urlprefix}{URL }
\expandafter\ifx\csname urlstyle\endcsname\relax
  \providecommand{\doi}[1]{doi:\discretionary{}{}{}#1}\else
  \providecommand{\doi}{doi:\discretionary{}{}{}\begingroup
  \urlstyle{rm}\Url}\fi
\providecommand{\eprint}[2][]{\url{#2}}

\bibitem{2017ApJ...848L..12A}
Abbott, B.~P., Abbott, R., Abbott, T.~D., Acernese, F., Ackley, K., et~al.
\newblock \emph{{Multi-messenger Observations of a Binary Neutron Star
  Merger}}.
\newblock \emph{Astrophysical Journal Letters}, \textbf{848(2)}:L12 (2017).
\newblock \doi{10.3847/2041-8213/aa91c9}.

\bibitem{2017ApJ...848L..13A}
Abbott, B.~P., Abbott, R., Abbott, T.~D., Acernese, F., Ackley, K., et~al.
\newblock \emph{{Gravitational Waves and Gamma-Rays from a Binary Neutron Star
  Merger: GW170817 and GRB 170817A}}.
\newblock \emph{Astrophysical Journal Letters}, \textbf{848(2)}:L13 (2017).
\newblock \doi{10.3847/2041-8213/aa920c}.

\bibitem{2019BAAS...51c..38B}
Burns, E., Tohuvavohu, A., Buckley, J., Canton, T.~D., Cenko, B., et~al.
\newblock \emph{{A Summary of Multimessenger Science with Neutron Star
  Mergers}}.
\newblock In \emph{BAAS}, volume~51, page~38 (2019).

\bibitem{IceCubeCollaboration2018a}
\emph{{Multimessenger observations of a flaring blazar coincident with
  high-energy neutrino IceCube-170922A}}.
\newblock \emph{Science}, \textbf{361(6398)}:eaat1378 (2018).
\newblock \doi{10.1126/science.aat1378}.
\newblock
  \urlprefix\url{http://science.sciencemag.org/content/361/6398/eaat1378.abstract}.

\bibitem{IceCubeCollaboration2018b}
IceCube~Collaboration, I.
\newblock \emph{{Neutrino emission from the direction of the blazar TXS
  0506+056 prior to the IceCube-170922A alert.}}
\newblock \emph{Science (New York, N.Y.)}, \textbf{361(6398)}:147--151 (2018).
\newblock ISSN 1095-9203.
\newblock \doi{10.1126/science.aat2890}.
\newblock \urlprefix\url{http://www.ncbi.nlm.nih.gov/pubmed/30002248}.

\bibitem{2016PhRvD..94j3006M}
Murase, K. and Waxman, E.
\newblock \emph{{Constraining high-energy cosmic neutrino sources: Implications
  and prospects}}.
\newblock \emph{Phys. Rev. D}, \textbf{94(10)}:103006 (2016).
\newblock \doi{10.1103/PhysRevD.94.103006}.

\bibitem{2017NatPh..13..232H}
Halzen, F.
\newblock \emph{{High-energy neutrino astrophysics}}.
\newblock \emph{Nature Physics}, \textbf{13(3)}:232--238 (2017).
\newblock \doi{10.1038/nphys3816}.

\bibitem{2019arXiv190610212M}
M{\'{e}}sz{\'{a}}ros, P., Fox, D.~B., Hanna, C., and Murase, K.
\newblock \emph{{Multi-Messenger Astrophysics}}.
\newblock \emph{arXiv e-prints}, page arXiv:1906.10212 (2019).

\bibitem{2019arXiv190304334A}
Ackermann, M., Ahlers, M., Anchordoqui, L., Bustamante, M., Connolly, A.,
  et~al.
\newblock \emph{{Astrophysics Uniquely Enabled by Observations of High-Energy
  Cosmic Neutrinos}}.
\newblock \emph{arXiv e-prints}, page arXiv:1903.04334 (2019).

\bibitem{2013PhRvD..88l1301M}
Murase, K., Ahlers, M., and Lacki, B.~C.
\newblock \emph{{Testing the hadronuclear origin of PeV neutrinos observed with
  IceCube}}.
\newblock \emph{Phys. Rev. D}, \textbf{88(12)}:121301 (2013).
\newblock \doi{10.1103/PhysRevD.88.121301}.

\bibitem{2018NatPh..14..396F}
Fang, K. and Murase, K.
\newblock \emph{{Linking high-energy cosmic particles by black-hole jets
  embedded in large-scale structures}}.
\newblock \emph{Nature Physics}, \textbf{14(4)}:396--398 (2018).
\newblock \doi{10.1038/s41567-017-0025-4}.

\bibitem{2019arXiv190304447B}
Buson, S., Fang, K., Keivani, A., Maccarone, T., Murase, K., et~al.
\newblock \emph{{A Unique Messenger to Probe Active Galactic Nuclei:
  High-Energy Neutrinos}}.
\newblock \emph{arXiv e-prints}, page arXiv:1903.04447 (2019).

\bibitem{2018ApJ...865..124M}
Murase, K., Oikonomou, F., and Petropoulou, M.
\newblock \emph{{Blazar Flares as an Origin of High-energy Cosmic Neutrinos?}}
\newblock \emph{ApJ}, \textbf{865(2)}:124 (2018).
\newblock \doi{10.3847/1538-4357/aada00}.

\bibitem{2019BAAS...51c..93S}
Sarazin, F., Anchordoqui, L., Beatty, J., Bergman, D., Covault, C., et~al.
\newblock \emph{{What is the nature and origin of the highest-energy particles
  in the universe?}}
\newblock In \emph{BAAS}, volume~51, page~93 (2019).

\bibitem{2019arXiv190306714A}
Alves~Batista, R., Biteau, J., Bustamante, M., Dolag, K., Engel, R., et~al.
\newblock \emph{{Open Questions in Cosmic-Ray Research at Ultrahigh Energies}}.
\newblock \emph{arXiv e-prints}, page arXiv:1903.06714 (2019).

\bibitem{1969PhLB...28..423B}
Beresinsky, V.~S. and Zatsepin, G.~T.
\newblock \emph{{Cosmic rays at ultra high energies (neutrino?)}}.
\newblock \emph{Physics Letters B}, \textbf{28(6)}:423--424 (1969).
\newblock \doi{10.1016/0370-2693(69)90341-4}.

\bibitem{Kotera2010}
Kotera, K., Allard, D., and Olinto, A.~V.
\newblock \emph{{Cosmogenic neutrinos: Parameter space and detectabilty from
  PeV to ZeV}}.
\newblock \emph{Journal of Cosmology and Astroparticle Physics},
  \textbf{2010(10)}:013--013 (2010).
\newblock ISSN 14757516.
\newblock \doi{10.1088/1475-7516/2010/10/013}.
\newblock
  \urlprefix\url{http://stacks.iop.org/1475-7516/2010/i=10/a=013?key=crossref.7fc7a70cb06490de9c0d9dada9148509}.

\bibitem{Kampert2012}
Kampert, K.~H. and Unger, M.
\newblock \emph{{Measurements of the cosmic ray composition with air shower
  experiments}}.
\newblock \emph{Astroparticle Physics}, \textbf{35(10)}:660--678 (2012).
\newblock ISSN 09276505.
\newblock \doi{10.1016/j.astropartphys.2012.02.004}.
\newblock
  \urlprefix\url{http://linkinghub.elsevier.com/retrieve/pii/S0927650512000382}.

\bibitem{ThePierreAugerCollaboration2016}
The Pierre Auger~Collaboration, A., Aab, A., Abreu, P., Aglietta, M., Ahn,
  E.~J., et~al.
\newblock \emph{{Evidence for a mixed mass composition at the `ankle' in the
  cosmic-ray spectrum}}.
\newblock \emph{Physics Letters B, Volume 762, p. 288-295.},
  \textbf{762}:288--295 (2016).
\newblock ISSN 0370-2693.
\newblock \doi{10.1016/j.physletb.2016.09.039}.
\newblock \urlprefix\url{http://dx.doi.org/10.1016/j.physletb.2016.09.039}.

\bibitem{Yushkov2017}
Yushkov, A.
\newblock \emph{{Recent results from the Pierre Auger Observatory on the mass
  composition and hadronic interactions of ultra-high energy cosmic rays}}.
\newblock In Pattison, B., editor, \emph{ISVHECRI 2016}, volume 145, page
  05002. EPJ Web of Conferences, Moscow (2017).
\newblock ISSN 2100-014X.
\newblock \doi{10.1051/epjconf/201614505002}.
\newblock
  \urlprefix\url{http://www.epj-conferences.org/10.1051/epjconf/201614505002}.

\bibitem{Aab2017a}
Aab, A., Abreu, P., Aglietta, M., Samarai, I.~A., Albuquerque, I., et~al.
\newblock \emph{{Combined fit of spectrum and composition data as measured by
  the Pierre Auger Observatory}}.
\newblock \emph{Journal of Cosmology and Astroparticle Physics},
  \textbf{2017(04)}:038--038 (2017).
\newblock ISSN 1475-7516.
\newblock \doi{10.1088/1475-7516/2017/04/038}.
\newblock
  \urlprefix\url{http://stacks.iop.org/1475-7516/2017/i=04/a=038?key=crossref.004396ad2440f77865500630b06473c9}.

\bibitem{Aab2017}
Aab, A., Abreu, P., Aglietta, M., Al~Samarai, I., Albuquerque, I. F.~M., et~al.
\newblock \emph{{Inferences on mass composition and tests of hadronic
  interactions from 0.3 to 100 EeV using the water-Cherenkov detectors of the
  Pierre Auger Observatory}}.
\newblock \emph{Physical Review D}, \textbf{96(12)}:122003 (2017).
\newblock ISSN 2470-0010.
\newblock \doi{10.1103/PhysRevD.96.122003}.
\newblock \urlprefix\url{https://link.aps.org/doi/10.1103/PhysRevD.96.122003}.

\bibitem{2019JCAP...01..002A}
Alves~Batista, R., de~Almeida, R.~M., Lago, B., and Kotera, K.
\newblock \emph{{Cosmogenic photon and neutrino fluxes in the Auger era}}.
\newblock \emph{Journal of Cosmology and Astroparticle Physics},
  \textbf{2019(1)}:2 (2019).
\newblock \doi{10.1088/1475-7516/2019/01/002}.

\bibitem{2019ApJ...872..110B}
Boncioli, D., Biehl, D., and Winter, W.
\newblock \emph{{On the Common Origin of Cosmic Rays across the Ankle and
  Diffuse Neutrinos at the Highest Energies from Low-luminosity Gamma-Ray
  Bursts}}.
\newblock \emph{Astrophysical Journal}, \textbf{872(1)}:110 (2019).
\newblock \doi{10.3847/1538-4357/aafda7}.

\bibitem{2018arXiv181210289Z}
Zhang, B.~T. and Murase, K.
\newblock \emph{{Ultrahigh-energy cosmic-ray nuclei and neutrinos from
  engine-driven supernovae}}.
\newblock \emph{arXiv e-prints}, page arXiv:1812.10289 (2018).

\bibitem{2005PhRvD..72b3001A}
Ahlers, M., Anchordoqui, L.~A., Goldberg, H., Halzen, F., Ringwald, A., et~al.
\newblock \emph{{Neutrinos as a diagnostic of cosmic ray galactic-extragalactic
  transition}}.
\newblock \emph{Phys. Rev. D}, \textbf{72(2)}:23001 (2005).
\newblock \doi{10.1103/PhysRevD.72.023001}.

\bibitem{2009APh....31..201T}
Takami, H., Murase, K., Nagataki, S., and Sato, K.
\newblock \emph{{Cosmogenic neutrinos as a probe of the transition from
  Galactic to extragalactic cosmic rays}}.
\newblock \emph{Astroparticle Physics}, \textbf{31(3)}:201--211 (2009).
\newblock \doi{10.1016/j.astropartphys.2009.01.006}.

\bibitem{2016PhRvD..94d3008L}
Liu, R.-Y., Taylor, A.~M., Wang, X.-Y., and Aharonian, F.~A.
\newblock \emph{{Indication of a local fog of subankle ultrahigh energy cosmic
  rays}}.
\newblock \emph{Phys. Rev. D}, \textbf{94(4)}:43008 (2016).
\newblock \doi{10.1103/PhysRevD.94.043008}.

\bibitem{2010PhRvD..81l3001M}
Murase, K. and Beacom, J.~F.
\newblock \emph{{Neutrino background flux from sources of ultrahigh-energy
  cosmic-ray nuclei}}.
\newblock \emph{Phys. Rev. D}, \textbf{81(12)}:123001 (2010).
\newblock \doi{10.1103/PhysRevD.81.123001}.

\bibitem{2019arXiv190101899V}
van Vliet, A., Alves~Batista, R., and H{\"{o}}rand~el, J.~R.
\newblock \emph{{Determining the fraction of cosmic-ray protons at ultra-high
  energies with cosmogenic neutrinos}}.
\newblock \emph{arXiv e-prints}, page arXiv:1901.01899 (2019).

\bibitem{GRANDCollaboration2018}
GRAND~Collaboration, G., Alvarez-Muniz, J., Batista, R.~A., V., A.~B., Bolmont,
  J., et~al.
\newblock \emph{{The Giant Radio Array for Neutrino Detection (GRAND): Science
  and Design}} (2018).
\newblock \urlprefix\url{http://arxiv.org/abs/1810.09994}.

\bibitem{2019arXiv190304333A}
Ackermann, M., Ahlers, M., Anchordoqui, L., Bustamante, M., Connolly, A.,
  et~al.
\newblock \emph{{Fundamental Physics with High-Energy Cosmic Neutrinos}}.
\newblock \emph{arXiv e-prints}, page arXiv:1903.04333 (2019).

\bibitem{2019arXiv190700991B}
Bhupal~Dev, P.~S., Babu, K.~S., Denton, P.~B., Machado, P. A.~N.,
  Arg{\"{u}}elles, C.~A., et~al.
\newblock \emph{{Neutrino Non-Standard Interactions: A Status Report}}.
\newblock \emph{arXiv e-prints}, page arXiv:1907.00991 (2019).

\bibitem{Cleveland1998}
Cleveland, B.~T., Daily, T., Davis, R., Jr., Distel, J.~R., Lande, K., et~al.
\newblock \emph{{Measurement of the Solar Electron Neutrino Flux with the
  Homestake Chlorine Detector}}.
\newblock \emph{The Astrophysical Journal}, \textbf{496(1)}:505--526 (1998).
\newblock ISSN 0004-637X.
\newblock \doi{10.1086/305343}.
\newblock \urlprefix\url{http://stacks.iop.org/0004-637X/496/i=1/a=505}.

\bibitem{Fukuda1998}
Fukuda, Y., Hayakawa, T., Ichihara, E., Inoue, K., Ishihara, K., et~al.
\newblock \emph{{Evidence for oscillation of atmospheric neutrinos}}.
\newblock \emph{Physical Review Letters}, \textbf{81(8)}:1562--1567 (1998).
\newblock ISSN 10797114.
\newblock \doi{10.1103/PhysRevLett.81.1562}.
\newblock \urlprefix\url{https://link.aps.org/doi/10.1103/PhysRevLett.81.1562}.

\bibitem{Ahmad2001}
Ahmad, Q.~R., Allen, R.~C., Andersen, T.~C., Anglin, J.~D., B{\"{u}}hler, G.,
  et~al.
\newblock \emph{{Measurement of the Rate of {$\nu$}e+d→p+p+e− Interactions
  Produced by 8B Solar Neutrinos at the Sudbury Neutrino Observatory}}.
\newblock \emph{Physical Review Letters}, \textbf{87(7)}:071301 (2001).
\newblock ISSN 0031-9007.
\newblock \doi{10.1103/PhysRevLett.87.071301}.
\newblock
  \urlprefix\url{https://link.aps.org/doi/10.1103/PhysRevLett.87.071301}.

\bibitem{Ahmad2002}
Ahmad, Q.~R., Allen, R.~C., Andersen, T.~C., D.Anglin, J., Barton, J.~C.,
  et~al.
\newblock \emph{{Direct Evidence for Neutrino Flavor Transformation from
  Neutral-Current Interactions in the Sudbury Neutrino Observatory}}.
\newblock \emph{Physical Review Letters}, \textbf{89(1)}:011301 (2002).
\newblock ISSN 10797114.
\newblock \doi{10.1103/PhysRevLett.89.011301}.
\newblock
  \urlprefix\url{https://link.aps.org/doi/10.1103/PhysRevLett.89.011301}.

\bibitem{IceCubeCollaboration2017}
IceCube~Collaboration, I., Aartsen, M.~G., Ackermann, M., Adams, J., Aguilar,
  J.~A., et~al.
\newblock \emph{{Measurement of the multi-TeV neutrino cross section with
  IceCube using Earth absorption}}.
\newblock \emph{Nature}, \textbf{551}:596--600 (2017).
\newblock \doi{10.1038/nature24459}.
\newblock \urlprefix\url{http://arxiv.org/abs/1711.08119}.

\bibitem{Aartsen2018a}
Aartsen, M., Ackermann, M., Adams, J., Aguilar, J., Ahlers, M., et~al.
\newblock \emph{{Measurement of Atmospheric Neutrino Oscillations at
  6–56 GeV with IceCube DeepCore}}.
\newblock \emph{Physical Review Letters}, \textbf{120(7)}:071801 (2018).
\newblock ISSN 0031-9007.
\newblock \doi{10.1103/PhysRevLett.120.071801}.
\newblock
  \urlprefix\url{https://link.aps.org/doi/10.1103/PhysRevLett.120.071801}.

\bibitem{Klein2013}
Klein, S.~R. and Connolly, A.
\newblock \emph{{Neutrino Absorption in the Earth, Neutrino Cross-Sections, and
  New Physics}} (2013).
\newblock \urlprefix\url{http://arxiv.org/abs/1304.4891}.

\bibitem{Ryabov2006}
Ryabov, V.~A.
\newblock \emph{{Ultrahigh-energy neutrinos from astrophysical sources and
  superheavy particle decays}}.
\newblock \emph{Physics-Uspekhi}, \textbf{49(9)}:905 (2006).
\newblock ISSN 1063-7869.
\newblock \doi{10.1070/PU2006v049n09ABEH006052}.
\newblock \urlprefix\url{http://stacks.iop.org/1063-7869/49/i=9/a=R02}.

\bibitem{2012JCAP...10..043M}
Murase, K. and Beacom, J.~F.
\newblock \emph{{Constraining very heavy dark matter using diffuse backgrounds
  of neutrinos and cascaded gamma rays}}.
\newblock \emph{Journal of Cosmology and Astroparticle Physics},
  \textbf{2012(10)}:43 (2012).
\newblock \doi{10.1088/1475-7516/2012/10/043}.

\bibitem{2018PhRvD..98h3016K}
Kachelrie{\ss}, M., Kalashev, O.~E., and Kuznetsov, M.~Y.
\newblock \emph{{Heavy decaying dark matter and IceCube high energy
  neutrinos}}.
\newblock \emph{Phys. Rev. D}, \textbf{98(8)}:83016 (2018).
\newblock \doi{10.1103/PhysRevD.98.083016}.

\bibitem{2019JCAP...05..051B}
Bhattacharya, A., Esmaili, A., Palomares-Ruiz, S., and Sarcevic, I.
\newblock \emph{{Update on decaying and annihilating heavy dark matter with the
  6-year IceCube HESE data}}.
\newblock \emph{Journal of Cosmology and Astroparticle Physics},
  \textbf{2019(5)}:51 (2019).
\newblock \doi{10.1088/1475-7516/2019/05/051}.

\bibitem{Gorham2018a}
Gorham, P.~W., Rotter, B., Allison, P., Banerjee, O., Batten, L., et~al.
\newblock \emph{{Observation of an Unusual Upward-Going Cosmic-Ray-like Event
  in the Third Flight of ANITA}}.
\newblock \emph{Physical Review Letters}, \textbf{121(16)}:161102 (2018).
\newblock ISSN 10797114.
\newblock \doi{10.1103/PhysRevLett.121.161102}.
\newblock
  \urlprefix\url{https://link.aps.org/doi/10.1103/PhysRevLett.121.161102}.

\bibitem{Gorham2019}
Gorham, P.~W., Allison, P., Banerjee, O., Batten, L., Beatty, J.~J., et~al.
\newblock \emph{{Constraints on the ultra-high energy cosmic neutrino flux from
  the fourth flight of ANITA}} (2019).
\newblock \urlprefix\url{http://arxiv.org/abs/1902.04005}.

\bibitem{2018arXiv180909615F}
Fox, D.~B., Sigurdsson, S., Shandera, S., M{\'{e}}sz{\'{a}}ros, P., Murase, K.,
  et~al.
\newblock \emph{{The ANITA Anomalous Events as Signatures of a Beyond Standard
  Model Particle, and Supporting Observations from IceCube}}.
\newblock \emph{arXiv e-prints}, page arXiv:1809.09615 (2018).

\bibitem{2019PhRvD..99d3009C}
Collins, J.~H., Dev, P.~S.~B., and Sui, Y.
\newblock \emph{{R -parity violating supersymmetric explanation of the
  anomalous events at ANITA}}.
\newblock \emph{Phys. Rev. D}, \textbf{99(4)}:43009 (2019).
\newblock \doi{10.1103/PhysRevD.99.043009}.

\bibitem{2019PhLB..790..578A}
Anchordoqui, L.~A. and Antoniadis, I.
\newblock \emph{{Supersymmetric sphaleron configurations as the origin of the
  perplexing ANITA events}}.
\newblock \emph{Physics Letters B}, \textbf{790}:578--582 (2019).
\newblock \doi{10.1016/j.physletb.2019.02.003}.

\bibitem{Otte2019a}
Otte, A.~N.
\newblock \emph{{Studies of an air-shower imaging system for the detection of
  ultrahigh-energy neutrinos}} (2019).
\newblock \doi{10.1103/PhysRevD.99.083012}.
\newblock \urlprefix\url{https://link.aps.org/doi/10.1103/PhysRevD.99.083012}.

\bibitem{Aharonian2006a}
Aharonian, F., Akhperjanian, A.~G., Bazer-Bachi, A.~R., Beilicke, M., Benbow,
  W., et~al.
\newblock \emph{{Observations of the Crab nebula with HESS}}.
\newblock \emph{Astronomy {\&} Astrophysics}, \textbf{457(3)}:899--915 (2006).
\newblock ISSN 0004-6361.
\newblock \doi{10.1051/0004-6361:20065351}.
\newblock \urlprefix\url{http://arxiv.org/abs/astro-ph/0607333
  http://dx.doi.org/10.1051/0004-6361:20065351
  http://www.aanda.org/10.1051/0004-6361:20065351}.

\bibitem{Aleksic2016}
Aleksi{\'{c}}, J., Ansoldi, S., Antonelli, L.~A., Antoranz, P., Babic, A.,
  et~al.
\newblock \emph{{The major upgrade of the MAGIC telescopes, Part II: A
  performance study using observations of the Crab Nebula}}.
\newblock \emph{Astroparticle Physics}, \textbf{72}:76--94 (2016).
\newblock ISSN 09276505.
\newblock \doi{10.1016/j.astropartphys.2015.02.005}.
\newblock
  \urlprefix\url{https://linkinghub.elsevier.com/retrieve/pii/S0927650515000316}.

\bibitem{Holder2016}
Holder, J. and Collaboration, f. t.~V.
\newblock \emph{{Latest Results from VERITAS: Gamma 2016}} (2016).
\newblock \urlprefix\url{http://arxiv.org/abs/1609.02881}.

\bibitem{ThePierreAugerCollaboration2017}
The Pierre Auger~Collaboration, T. P.~A., Aab, A., Abreu, P., Aglietta, M.,
  Albuquerque, I. F.~M., et~al.
\newblock \emph{{The Pierre Auger Observatory: Contributions to the 35th
  International Cosmic Ray Conference (ICRC 2017)}} (2017).
\newblock \urlprefix\url{http://arxiv.org/abs/1708.06592}.

\bibitem{Tokuno2012}
Tokuno, H., Tameda, Y., Takeda, M., Kadota, K., Ikeda, D., et~al.
\newblock \emph{{New air fluorescence detectors employed in the Telescope Array
  experiment}}.
\newblock \emph{Nuclear Instruments and Methods in Physics Research Section A:
  Accelerators, Spectrometers, Detectors and Associated Equipment},
  \textbf{676}:54--65 (2012).
\newblock ISSN 0168-9002.
\newblock \doi{10.1016/J.NIMA.2012.02.044}.
\newblock
  \urlprefix\url{https://www.sciencedirect.com/science/article/pii/S0168900212002422?via%3Dihub}.

\bibitem{Fargion1999}
Fargion, D., Aiello, A., and Conversano, R.
\newblock \emph{{Horizontal Tau air showers from mountains in deep valley.
  Traces of UHECR neutrino tau}} (1999).
\newblock \urlprefix\url{http://arxiv.org/abs/astro-ph/9906450}.

\bibitem{Ahnen2018a}
Ahnen, M.~L., Ansoldi, S., Antonelli, L.~A., Arcaro, C., Baack, D., et~al.
\newblock \emph{{Limits on the flux of tau neutrinos from 1 PeV to 3 EeV with
  the MAGIC telescopes}}.
\newblock \emph{Astroparticle Physics}, \textbf{102}:77--88 (2018).
\newblock ISSN 09276505.
\newblock \doi{10.1016/j.astropartphys.2018.05.002}.
\newblock
  \urlprefix\url{https://www.sciencedirect.com/science/article/pii/S0927650518300136?via%3Dihub}.

\bibitem{Anderhub2013}
Anderhub, H., Backes, M., Biland, A., Boccone, V., Braun, I., et~al.
\newblock \emph{{Design and operation of FACT-the first G-APD Cherenkov
  telescope}}.
\newblock \emph{Journal of Instrumentation}, \textbf{8(6)}:P06008--P06008
  (2013).
\newblock ISSN 17480221.
\newblock \doi{10.1088/1748-0221/8/06/P06008}.
\newblock
  \urlprefix\url{http://stacks.iop.org/1748-0221/8/i=06/a=P06008?key=crossref.215223ca70483ee275407a04fbb44da8}.

\bibitem{Cortina2016}
Cortina, J., L{\'{o}}pez-Coto, R., and Moralejo, A.
\newblock \emph{{MACHETE: A transit imaging atmospheric Cherenkov telescope to
  survey half of the very high energy {$\gamma$}-ray sky}}.
\newblock \emph{Astroparticle Physics}, \textbf{72}:46--54 (2016).
\newblock ISSN 09276505.
\newblock \doi{10.1016/j.astropartphys.2015.07.001}.
\newblock \urlprefix\url{http://arxiv.org/abs/1507.02532
  http://dx.doi.org/10.1016/j.astropartphys.2015.07.001}.

\bibitem{Gomez2016}
G{\'{o}}mez, S., Gasc{\'{o}}n, D., Fern{\'{a}}ndez, G., Sanuy, A., Mauricio,
  J., et~al.
\newblock \emph{{MUSIC: An 8 channel readout ASIC for SiPM arrays}}.
\newblock volume 9899, page 98990G. International Society for Optics and
  Photonics (2016).
\newblock ISBN 978-1-5106-0144-4.
\newblock ISSN 0277-786X.
\newblock \doi{10.1117/12.2231095}.
\newblock
  \urlprefix\url{http://proceedings.spiedigitallibrary.org/proceeding.aspx?doi=10.1117/12.2231095}.

\bibitem{Pollacco2018}
Pollacco, E., Grinyer, G., Abu-Nimeh, F., Ahn, T., Anvar, S., et~al.
\newblock \emph{{GET: A generic electronics system for TPCs and nuclear physics
  instrumentation}}.
\newblock \emph{Nuclear Instruments and Methods in Physics Research Section A:
  Accelerators, Spectrometers, Detectors and Associated Equipment},
  \textbf{887}:81--93 (2018).
\newblock ISSN 01689002.
\newblock \doi{10.1016/j.nima.2018.01.020}.
\newblock
  \urlprefix\url{https://www.sciencedirect.com/science/article/pii/S0168900218300342}.

\bibitem{Otte2019b}
Otte, A.~N.
\newblock \emph{{Development of a Cherenkov Telescope for the Detection of
  Ultra-High Energy Neutrinos with EUSO-SPB2 and POEMMA}}.
\newblock In \emph{36th International Cosmic Ray Conference}, page~8.
  Proceedings of Science, Madison, WI (2019).

\bibitem{brown2015}
Brown, A.~M., Armstrong, T., Chadwick, P.~M., Daniel, M., White, R., et~al.
\newblock \emph{{Flasher and muon-based calibration of the GCT telescopes
  proposed for the Cherenkov Telescope Array}}.
\newblock \emph{arXiv e-prints} (2015).

\bibitem{2018APh....97...69B}
Brown, A.~M.
\newblock \emph{{On the prospects of cross-calibrating the Cherenkov Telescope
  Array with an airborne calibration platform}}.
\newblock \emph{Astroparticle Physics}, \textbf{97}:69--79 (2018).
\newblock \doi{10.1016/j.astropartphys.2017.10.013}.

\end{thebibliography}

\end{document}